# Threshold Values of Random $K$-SAT from the Cavity Method

Stephan Mertens [1]

*Institut für Theoretische Physik, Otto-von-Guericke Universität, Postfach 4120, 39016 Magdeburg, Germany*

Marc Mézard [2]

*CNRS, Laboratoire de Physique Théorique et Modèles Statistiques, Université Paris Sud, 91405 Orsay Cedex, France*

Riccardo Zecchina [3]

*The Abdus Salam International Centre for Theoretical Physics, St. Costiera 11, 34100 Trieste, Italy*

**Abstract**

Using the cavity equations of [23,24], we derive the various threshold values for the number of clauses per variable of the random $K$-satisfiability problem, generalizing the previous results to $K \geq 4$. We also give an analytic solution of the equations, and some closed expressions for these thresholds, in an expansion around large $K$. The stability of the solution is also computed. For any $K$, the satisfiability threshold is found to be in the stable region of the solution, which adds further credit to the conjecture that this computation gives the exact satisfiability threshold.

*Key words:* Satisfiability; K-SAT; Threshold Phenomenon; Phase Transition; Cavity Approach; Survey Propagation; Average Case Complexity

## 1 Introduction

The $K$-satisfiability problem (K-SAT) is easily stated: Given $N$ Boolean variables each of which can be assigned the value True or False, and $M$ constraints between them taking the form of clauses, is there a 'SAT-assignment', i.e. an assignment of the Boolean variables which satisfies all constraints. A clause takes the form of an 'OR' function of $K$ variables in the ensemble (or their negations).

[1] stephan.mertens@physik.uni-magdeburg.de
[2] mezard@lptms.u-psud.fr
[3] zecchina@ictp.trieste.it

K-SAT plays a central role in computer science, and a lot of efforts have been devoted to this problem. As soon as there are clauses with $K \geq 3$ variables this problem is NP-complete [7,13]. In recent years, the interest has focused on the random K-SAT problem: all the clauses involve the same number $K$ of variables; the variables in each clause are chosen at random uniformly in the set of all variables, and a variable appears negated with probability 1/2. This problem displays a very interesting threshold phenomenon when one takes the large $N$ limit, keeping the ratio of clauses to variable, $\alpha = M/N$, fixed. Numerical simulations [16] suggest the existence of a phase transition at a value $\alpha_c(K)$ of this ratio: For $\alpha < \alpha_c(K)$ a randomly generated problem is satisfiable (SAT) with probability going to one in the large $N$ limit, for $\alpha > \alpha_c(K)$ a randomly generated problem is not satisfiable (UNSAT) with probability going to one in the large $N$ limit. This phase transition is particularly interesting because it turns out that the really difficult instances, from the algorithmic point of view, are those where $\alpha$ is close to $\alpha_c$. The study of this phase transition is thus one step in the elaboration of a theory of typical case complexity, [19,33,35] and it has attracted a lot of interest [15,6,9,32].

On the analytical side, there exists a proof that the threshold phenomenon exists at large $N$[11], although the fact that the corresponding $\alpha_c$ has a limit when $N \to \infty$ has not yet been established rigorously. Upper bounds $\alpha_{\text{UB}}(K)$ on $\alpha_c$ have been found using first moment methods [8,17] and variational interpolation methods [14,10], and lower bounds $\alpha_{\text{LB}}(K)$ have been found using either explicit analysis of some algorithms[1], or some second moment methods [2].

Recently, some of the powerful concepts and techniques of statistical physics [21] have been applied to this problem [9,25,26,23,24], focusing on the case of $K = 3$. Using a heuristic method called in physics jargon the 'one step replica symmetry breaking' (1RSB) cavity method, the threshold has been conjectured to be $\alpha_c(K = 3) \simeq 4.267$ [23,24].

The present paper presents a reformulation of the 1RSB cavity method computation of [23,24], in which no physics background is assumed. The method can be understood as a pure statistical analysis of the properties of survey propagation equations of [24,4]. These are message passing equations, defined on the factor graph representation of the satisfiability problem, which provide a kind of generalization of the well known belief propagation equations. Starting from these equations, we show the various steps of the statistical analysis which give, for any $K$, the satisfiability threshold $\alpha_c(K)$. The results are sum-marized in table 1 in sec. 5.3. For general $K$ these computations involve some complicated integral equations, which can be solved numerically for $K$ not too large. We also study in details the large $K$ limit, where we can compute the solution analytically and work out a series expansion of the satisfiability threshold in powers of $2^{-K}$.

A second aspect of the paper is a check of self-consistency of the cavity approach: this is a heuristic approach which relies on some hypothesis of absence of correlations between some random variables. While we cannot prove the validity of this hypothesis, it is possible to check whether it is self-consistent, using a method recently developed by Montanari and Ricci-Tersenghi [28]. The 1-RSB cavity solution is found to be self-consistent in a finite window of $\alpha$ below the satisfiability threshold, $\alpha_s(K) < \alpha \leq \alpha_c(K)$. For $K = 3$ we find $\alpha_s(3) \simeq 4.15$ and results for larger $K$ are also provided in table 1. For all $K$, we find that the cavity 1RSB analysis is self-consistent at $\alpha_c(K)$. We can thus conjecture that the values of $\alpha_c(K)$ which we give here are the exact values for the threshold.

The type of approach which we use here is heuristic. The self-consistency of its assumptions can be checked, and its predictions can be tested against numerical simulations. Eventually one may hope to gain enough insight from this approach in order to convert it into a full proof. It is interesting to notice that in the simpler case of the random K-XORSAT problem, where the same kind of phase transition has been found [30], the full structure and the results of the cavity 1RSB solution can be confirmed rigorously [22]. Recently, the application to random K-SAT of the variational interpolation method of Guerra [14] has allowed to prove that for even K the cavity result for $\alpha_c(K)$ is an upper bound to the true threshold values[10].

Sect. 2 summarizes the survey propagation equations which are at the heart of the cavity method; it also provides a short account of the intuitive interpretation in terms of clusters of SAT assignements. The statistical analysis of these equations is described in sect. 3. In sect. 4 we explain how to compute the stability of this solution. Sect. 5 contains a numerical analysis of the basic equations using a population dynamics method. Threshold values are computed for various values of $K$. Sect. 6 provides an analytic study of the basic equations, using a large $K$ expansion. It is used to compute the large $K$ behavior of the various thresholds, and in particular it derives a series expansion for the satisfiability threshold, which matches the numerical results of sect. 5. Conclusions are summarized in sect 7: There we have formulated our main results in the form of a series of explicit conjectures. An appendix contains some details of the large $K$ series expansion of the satisfiability threshold.

## 2 Background

### 2.1 Survey propagation equations

The 'survey-propagation' (SP) heuristic, which has been described in details in [24,4], is an iterative message passing algorithm which turns out to be effective in finding SAT assignments in the $\alpha < \alpha_c$ region, quite close to the threshold. Here we briefly recall the explicit form of the SP equations which are the basis for the algorithm; a more detailed (still heuristic) derivation can be found in [4].

An instance of $K$-SAT can be represented by a bipartite graph, the so called *factor graph* [18]. Each clause corresponds to a function node, each variable to a variable node, and an edge connects a function node and a variable node if and only if the clause contains the variable. Fig. 1 shows part of a factor graph, with clauses (function nodes) denoted by squares and variable nodes denoted by circles. A dashed edge between a clause and a variable means that the variable appears negated in the clause, a full edge means that it appears unnegated. Associated to each edge $a-f$ there is a real number $\eta_{a \to f} \in [0,1]$, called cavity-bias survey, which is the message sent from clause $a$ to variable $f$. This message is computed from the messages received by the $K-1$ 'input' variables $i$ which are involved in clause $a$, but distinct from $f$. Consider the corresponding portion of the factor graph as shown in Fig. 1. For any 'input' variable $i$, we define:

$$\pi_\pm^i = \prod_{b \in V_\pm^{(i)}} (1 - \eta_{b \to i}) \qquad (1)$$

where $V_+^{(i)}$ denotes the set of all function nodes (except $a$, the 'cavity') in which variable $i$ appears unnegated and $V_-^{(i)}$ denotes the corresponding set of clauses where $i$ appears negated. Then the 'output' cavity-bias survey $\eta_{a \to f}$ is given by

$$\eta_{a \to f} = \prod_{i=1}^{K-1} \frac{\pi_+^i (1 - \pi_-^i)}{\pi_+^i + \pi_-^i - \pi_+^i \pi_-^i} \, . \qquad (2)$$

In order to keep notations simple, this equation is written for the situation of Fig. 1; if the edge between $a$ and $f$ were a dashed edge, the roles of $\pi_+^i$ and $\pi_-^i$ should be inverted in (2). Eqs. (1,2) are the SP equations [4].

These equations have a simple interpretation in terms of 'warnings' [4]. A warning sent from clause $a$ to variable $f$ indicates that $f$ should be assigned the value that satisfies $a$, and $a$ will send this warning if and only if all of its other variables are constrained to non-satisfying values. For instance in Fig. 1 a warning travels from $a$ to $f$ only if each of the variables $1, \ldots, K-1$ receives an 'impeding warning', that is a warning which impedes it to satisfy $a$. The following reasoning shows that $\eta_{a \to f}$ can be interpreted as the probability that the function node sends a warning to the variable node $f$, assuming that the incoming warnings arriving on the variables $i = 1, ..K - 1$ are independent random variables. A variable $i$ can receive either an impeding warning (coming from one of its neighbours in $V_-^{(i)}$) telling it to take the value which does not satisfy $a$, or a 'supporting warning' (coming from one of its neighbours in $V_+^{(i)}$) telling it to take the value which satisfies $a$. From (1), it is clear that $\pi_+^i$ denotes the probability that variable $i$ receives no supporting warning, and $\pi_-^i$ denotes the probability that variable $i$ receives no impeding warning. So there are 4 possible cases for variable $i$:

- It receives no warning at all: probability $\pi_+^i \pi_-^i$.
- It receives at least one impeding warning, but no supporting warning: probability $\pi_+^i (1 - \pi_-^i)$.
- It receives at least one supporting warning, but no impeding warning: probability $\pi_-^i (1 - \pi_+^i)$.
- It receives at least one impeding warning and at least one supporting warning: probability $(1 - \pi_+^i)(1 - \pi_-^i)$.

The last case implies that there is no satisfying assignment. Since we restrict ourselves to satisfying assignments we have to condition the probabilities on the first three cases only. Then the probability that $i$ receives at least one impeding warning, given that there are no contradictions, is $\pi_+^i (1 - \pi_-^i) / (\pi_+^i + \pi_-^i - \pi_+^i \pi_-^i)$. Clause $a$ will send a warning to $f$ if and only if all its input neighbours $i$ are in this situation: this gives equation (2).

This shows that the SP equations are exact whenever the incoming warnings are independent random variables. This is the case in particular if the factor graph is a tree. In the random K-SAT problem the factor graph is not a tree, but a random bipartite graph, with fixed degree (equal to $K$) for function nodes, and Poisson distributed degrees (as $N \to \infty$) for variable nodes. This graph is locally tree-like in the following sense: For any fixed $r$, if one picks up one vertex at random and considers its neighborhood up to distance $r$, this neighborhood is a tree with probability going to one in the large $N$ limit. Therefore one may hope that the survey propagation equations are relevant for the description of the random K-SAT problem, although there does not exist any proof of this statement.

2.2 *Physical interpretation: clusters of SAT assignments*

As we have seen the cavity-bias surveys have all mathematical properties of probabilities of warnings. An obvious question is: what is the corresponding probability space? The statistical physics heuristic 'derivation' of SP equations

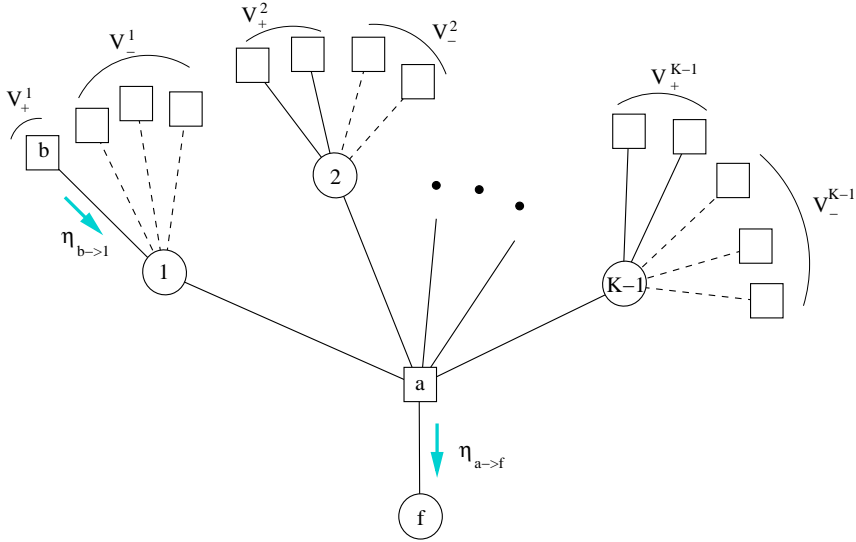

Fig. 1. Part of a factor graph representing a satisfiability problem. Clauses are denoted by squares, variables by circles. A dashed line between a variable and a clause means that the variable appears negated in the clause, a full line means that it appears unnegated. The clause $a$ is connected to the variables $1, ..., K-1, f$. The cavity-bias-survey $\eta_{a \to f}$ sent from clause $a$ to variable $f$ depends on the set of all cavity-bias-surveys like $\eta_{b \to i}$ which arrive onto variables $i \in \{1, ..., K-1\}$ from the other clauses (distinct from $a$). Its explicit expression is given in (2)

suggests a conjecture for this probability space, which we now briefly describe. For the physics reasoning that leads to this conjecture we refer the interested reader to [20,24].

We focus on satisfiable instances. For each such instance, with $N$ variables, consider the set of SAT assignments, which is a subset of the unit hypercube. We use the Hamming distance between two assignments defined as the number of variables in which they differ. The first assumption concerns the topology of the set of SAT assignments, which is supposed to break into clusters. We first introduce the definitions of clusters and constrained clusters:

A **$d$-cluster** is a subset of SAT-assignments obtained as follows: build an auxiliary graph where every SAT-assignment is a vertex, and one puts an edge between two SAT-assignment if and only they are at distance $\leq d$. Every connected component of this auxiliary graph is a $d$-cluster of SAT configurations.

If there exists a variable which takes the same value in *all* the SAT-assignments of a given $d$-cluster, this cluster is called a **constrained $d$-cluster**.

In the type of cavity solution which has been worked out so far, called 'one step replica symmetry breaking' (1-RSB) in the statistical physics jargon, it is assumed that, in a window $\alpha_s < \alpha < \alpha_c$, there exist, in the large $N$ limit, exponentially many well separated constrained $d$-clusters. Unfortunately it is not known what value of $d$ should be used in this analysis. A common belief is that $d(N)$ may grow with $N$, but with $\lim_{N\to\infty} d(N)/N = 0$. We shall loosely use the word 'clusters' instead of '$d$-clusters'. Note that in the analysis developped in this paper the limit $N \to \infty$ has been taken implicitly and from the very beginning. Hence we do not need to specify $d(N)$.

2.3 *Complexity*

An important quantity is the number $\mathcal{N}_c$ of constrained clusters. It is assumed that, when one generates random instances of K-SAT, the distribution of $(1/N)\ln(\mathcal{N}_c)$ becomes sharply concentrated in the large $N$ limit, and one defines the **complexity** as $\Sigma = \lim_{N\to\infty}(1/N)\mathbb{E}(\ln \mathcal{N}_c)$. Furthermore the constrained clusters are supposed to be well separated at large $N$ in the following sense: Taking two SAT-assignments randomly chosen inside the same constrained cluster, the distribution of $1/N$ times their distance is assumed to be sharply concentrated around a value $d_1$ which measures the size of the constrained cluster in the sense that $d_1$ is the typical distance between solutions randomly chosen from the same cluster. For two SAT-assignments randomly chosen inside different constrained clusters, this same quantity is peaked around $d_0$, the inter-cluster distance, and $d_0 > d_1$.

Statistical physics techniques are able to check the self-consistency of these assumptions, by some kind of circular argument, and within this framework they provide some educated conjectures concerning the values of $\alpha_s$ and $\alpha_c$, as we shall see below. But they cannot prove the existence of clusters. We thus leave the clustering property of the space of satisfying assignments as a mathematical conjecture.

Assuming that constrained clusters exist for some $\alpha$, it is useful to introduce a generalized space for variables, where each variable can take three values : 0 for a variable constrained to FALSE, 1 for a variable constrained to TRUE, $*$ for an unconstrained variable. A constrained cluster is then characterized by a single point (different from the 'all $*$' point) in the generalized variable space $\{0, 1, *\}^N$. One can associate to the constrained cluster one set of warnings. The cavity-bias survey $\eta_{a\to f}$ is interpreted as the probability that a warning is present on the edge $a$ to $f$ when one picks up a constrained cluster at random. It has been argued in [4], and shown in [5], that the SP equations can be interpreted as standard belief propagation (BP) equations in the generalized variable space $\{0, 1, *\}^N$. Because BP allows to count the number of solutions, it gives an explicit formula to obtain the number of constrained clusters. The final result for the complexity $(1/N)\mathbb{E}(\ln \mathcal{N}_c)$ for a given instance with $N \gg 1$ is (we refer the reader to [4] for a detailed explanation of this computation):

$$\Sigma = \frac{1}{N} \left[ \sum_{a=1}^{M} \Sigma_a^c - \sum_{j=1}^{N} (n_j - 1) \Sigma_j^v \right] \quad (3)$$

where $\Sigma_a^c$ is the contribution from clause $a$ and $\Sigma_j^v$ is the contribution from variable $j$, $n_j$ being the degree of this variable. Keeping the notations of Fig. 1, and considering $f$ as the $K$-th neighbour of clause $a$, the contribution of this clause is:

$$\Sigma_a^c = \ln \left[ \prod_{i=1}^{K} \left( \pi_+^i + \pi_-^i - \pi_+^i \pi_-^i \right) - \prod_{i=1}^{K} \left( \pi_+^i (1 - \pi_-^i) \right) \right] . \quad (4)$$

The contribution of site $j$ is

$$\Sigma_j^v = \ln \left[ \prod_{b \in W_+^{(j)}} (1 - \eta_{b \to j}) + \prod_{b \in W_-^{(j)}} (1 - \eta_{b \to j}) - \prod_{b \in W^{(j)}} (1 - \eta_{b \to j}) \right] \quad (5)$$

Where $W_+^{(j)}$ is the set of function nodes connected to $j$ by a full edge, $W_-^{(j)}$ is the set of function nodes connected to $j$ by a dashed edge, and $W^{(j)} = W_+^{(j)} \cup W_-^{(j)}$.

In practice, explicit computations of the complexity performed below show that it is is a decreasing function of $\alpha$ which vanishes at some value. It is then a natural conjecture, totally unproven so far, that the satisfiability threshold $\alpha_c(K)$ is equal the value of $\alpha$ where the complexity $\Sigma$ vanishes. This is the criterion which we shall use in the rest of the paper.

## 3 Statistical analysis

In this section we show how the statistical analysis of the SP equations (1,2), and of the complexity (3), leads to coupled integral equations from which one can compute the thresholds. Such equations were first derived for $K = 3$ in [24]; we provide here a simpler derivation, restricted to satisfying assignments, but valid for any $K$.

Picking an instance of random K-SAT, and a random edge in the corresponding factor graph, the cavity-bias survey $\eta$ on this edge is a random variable. We would like to compute the probability distribution of this random variable. Because of the local tree-like structure of the factor graph, the variables $\pi_\pm^i$ appearing in (2) are assumed independent. To compute their distribution, we first notice that $\pi_+^i$ is a product of $k_+^i$ factors $1 - \eta$, where $k_+^i$ is the number of clauses $b \in V_+^i$ (see Fig. 1), conditioned to the fact that there exists an edge $i - a$. Let us study its distribution. We first notice that, in the large $N$ limit, the degree of a variable is Poisson distributed with mean $K\alpha$. Also, the total number of full edges (resp. of dashed edges) ending on a variable are two iid Poisson distributed variables with mean $K\alpha/2$. As the presence of the edge $i - a$ is an event independent from the presence of other edges, it turns out that $k_+^i$ and $k_-^i$ also are iid variables with a Poisson distribution of mean $K\alpha/2$.

It is useful to change variables, introducing for each survey $\eta$ and for each factor $\pi$ the variables

$$\phi = -\ln(1 - \eta) \text{ and } x = -\ln(\pi). \tag{6}$$

We call $S(\phi)$ the probability density function (pdf) of $\phi$ and $B(x)$ the pdf of $x$. These are two positive functions on $[0, \infty[$ with integral equal to one. Note that both pdf's are *mixtures* in the sense that they are atomic at argument 0 (see sec. 5.1).

For each $k$, the distribution of $x$ is the $k$−th convolution, $S^{\otimes k}$ of $S$. Summing over $k$ we get:

$$B(x) = \sum_{k=0}^{\infty} f_{K\alpha/2}(k) S^{\otimes k}(x) \tag{7}$$

where $f_{K\alpha/2}(k) = \frac{(K\alpha/2)^k}{k!} \exp(-K\alpha/2)$ is the Poisson probability mass function for the connectivity. Using the same indices as in Fig. 1 for the new variables, the SP equation (2) becomes:

$$\phi_{a \to f} = -\ln \left[ 1 - \prod_{i=1}^{K-1} \frac{e^{x_-^i} - 1}{e^{x_+^i} + e^{x_-^i} - 1} \right]. \tag{8}$$

As the variables $x_\pm^i$ are iid with pdf $B$, one obtains from this equation the pdf of $\phi_{a \to f}$. Identifying it with $S$ shows that this pdf $S(\phi)$ satisfies the equation:

$$S(\phi) = \int \prod_{i=1}^{K-1} \left[ B(x_+^i) \mathrm{d}x_+^i \ B(x_-^i) \mathrm{d}x_-^i \right] \delta \left( \phi + \ln \left[ 1 - \prod_{i=1}^{K-1} \frac{e^{x_-^i} - 1}{e^{x_+^i} + e^{x_-^i} - 1} \right] \right) \tag{9}$$

Equations (7) and (9) provide two coupled equations for the pdfs $S(\phi)$ and $B(x)$.

Once these have been determined the complexity $\Sigma$ can be computed from (3), (4) and (5) as follows. We first write (3) as:

$$\Sigma = \frac{1}{N} \left[ \sum_{a=1}^{M} \Sigma_a^c + \sum_{j=1}^{N} \Sigma_j^v - \sum_{a=1}^{M} \sum_{j \in V(a)} \Sigma_{aj}^{cv} \right], \tag{10}$$

where $\Sigma_a^c$ and $\Sigma_j^v$ have been defined in (4,5), and the $\Sigma_{aj}^{cv}$ for a full edge is

written as:

$$\Sigma_{aj}^{cv} = \ln\left[(1-\eta_{a\to j})\prod_{b\in V_+^{(j)}\setminus a}(1-\eta_{b\to j}) + \prod_{b\in V_-^{(j)}}(1-\eta_{b\to j})\right.$$
$$\left. - (1-\eta_{a\to j})\prod_{b\in V^{(j)}\setminus a}(1-\eta_{b\to j})\right] \quad (11)$$

the one for a dashed edge being written similarly. One then computes the expectation value of each term in (10). Using the change of variables (6), one obtains:

$$\mathbb{E}\left(\Sigma_a^c\right) = \mathbb{E}\left(\ln\left[\prod_{i=1}^K\left(e^{-x_i}+e^{-y_i}-e^{-(x_i+y_i)}\right) - \prod_{i=1}^K\left(e^{-x_i}-e^{-(x_i+y_i)}\right)\right]\right), \quad (12)$$

where the expectation value $\mathbb{E}()$ refers to an average where all the variables $x_i$ and $y_i$ are drawn from $B(x)$. This in turn gives $\mathbb{E}\left(\Sigma_a^c\right) = -2K\mathbb{E}(x_1)+I_K$, where we define:

$$I_s = \int\prod_{i=1}^s \mathrm{d}x_i\mathrm{d}y_i\, B(x_i)B(z_i)\ln\left[\prod_{i=1}^s(\mathrm{e}^{x_i}+\mathrm{e}^{z_i}-1) - \prod_{i=1}^s(\mathrm{e}^{x_i}-1)\right]. \quad (13)$$

Similarly, using the fact that

$$x = -\ln\left[\prod_{b\in V_+^{(j)}}(1-\eta_{b\to j})\right]$$
$$z = -\ln\left[\prod_{b\in V_-^{(j)}}(1-\eta_{b\to j})\right] \quad (14)$$

are both distributed with $B$, one finds $\mathbb{E}\left(\Sigma_j^v\right) = -2\mathbb{E}(x)+\Sigma_0$, where

$$\Sigma_0 = \int \mathrm{d}x\mathrm{d}z\, B(x)B(z)\ln(\mathrm{e}^x+\mathrm{e}^z-1). \quad (15)$$

Considering $\Sigma_{aj}^{cv}$, it can be written as:

$$\Sigma_{aj}^{cv} = +\ln\left[(1-\eta_{a\to j})e^{-x_K}+e^{-y_K}-(1-\eta_{a\to j})e^{-(x_K+y_K)}\right], \quad (16)$$

where $x_K$ and $y_K$ are two random variables with pdf $B$. Using the equations (6,8), the term $1-\eta_{a\to j}$ can be written in terms of $2(K-1)$ random variables, $x_i, y_i, i \in \{1,\ldots,K-1\}$, all with pdf $B$. Substituting this into (16), one gets $\mathbb{E}\left(\Sigma_{aj}^{cv}\right) = -2\mathbb{E}(x)-(K-1)\Sigma_0+I_K$. A further simplification can be obtained by noticing that $\mathbb{E}(x) = \frac{K\alpha}{2}\mathbb{E}(\Phi)$ (Eq. 7) and $I_{K-1} = (K-1)\Sigma_0-\mathbb{E}(\Phi)$ which allow us to write

$$\mathbb{E}(x) = \frac{K\alpha}{2}[(K-1)\Sigma_0-I_{K-1}] \quad (17)$$

Putting all this together we finally obtain

$$\Sigma = \Sigma_0 + \alpha[KI_{K-1} - (K-1)I_K], \qquad (18)$$

where $\Sigma_0$ and $I_s$ are defined in (15,13). Once the distribution $B(x)$ has been determined, the complexity can be easily computed. The rest of the paper is devoted to the calculation of $S(\phi)$, $B(x)$ and $\Sigma$, but first we discuss the stability of the underlying 1-RSB hypothesis.

## 4 Stability analysis

### 4.1 General formalism

The above SP equations have been derived within the hypothesis of a 'one step RSB' hypothesis. The stability of a this type of solution with respect to two steps RSB has been discussed in details by Montanari and Ricci-Tersenghi [28]. We shall apply their method to $K$-satisfiability, using the presentation developed in [27,31], which lends itself to a direct interpretation in terms of the stability of the message passing procedure. Recently the same analysis has been done independently for $K = 3$ and $K = 4$ in [29].

We first explain this general formalism. The survey propagation equation (2) can be written in a very general form as:

$$p_\gamma^\ell = \frac{1}{Z} \sum_{(\beta_1,...,\beta_n)\to\gamma} p_{\beta_1}^1....p_{\beta_n}^n \chi(\beta_1,...,\beta_n), \qquad (19)$$

where the indices $\gamma, \beta_i$ refer to some type of warnings which can be transmitted on a link. $p_\gamma^\ell$ is the probability of having a warning of type $\gamma$ on a link with index $\ell$; it depends on the warnings $\beta_1,...,\beta_n$ being sent on $n$ other links, and $\chi$ is a general function of these $n$ warnings. In the case of the satisfiability problem, on a given link $\ell = a \to f$ from a function node $a$ to a variable node $f$ there are only two possible elementary messages: warning ($\gamma = 1$) or no warning ($\gamma = 0$). Referring to figure 1, we have thus $p_{\gamma=1}^\ell = \eta_{a\to f}$ which depends on the various $\eta_{b\to i}$ numbers, and $\chi$ is the indicator which is equal to zero if there is a contradiction in the incoming messages, equal to 1 otherwise, and $Z$ is a normalization factor.

The propagation equation (19) can have two types of instabilities[28], which correspond to the two ways a one step RSB solution can be unstable with respect to two steps of RSB, but which also have a direct interpretation.

- A first type of instability, called of type I in the nomenclature of [28],

amounts to see if a small change of one probability propagates. This iteration stability is computed from the study of the Jacobian

$$T_{\gamma\beta} = \frac{\partial p_\gamma^\ell}{\partial p_\beta^1} \tag{20}$$

of equation (19). This matrix describes the propagation of a perturbation after one iteration. Following the perturbation by iterating $d$ times the SP equations, one gets a product of $d$ such Jacobian matrices, $T^1...T^d$, each of them being different since the values of the surveys vary from one link to the next. The global perturbation induced after $d$ iterations by a change $p_b^1 \to p_b^1 + dp_b^1$ concerns on average $(K\alpha(K-1))^d$ cavity-biases ($K\alpha$ is the average connectivity of a variable, $K-1$ is the number of branches along which the perturbation propagates when encountering a function node). The perturbation is monitored by the sum of the squares of the perturbed cavity-biases, which behaves like

$$[K\alpha(K-1)]^d \ \mathrm{Tr}\langle[T^1...T^d]^2\rangle \equiv \lambda_d \ . \tag{21}$$

where $\langle\rangle$ means an average over all possible cavity bias surveys. By taking the squares of the perturbed cavity in (21) we cover both signs of the perturbation. In general, one finds that $\lambda_d$ depends exponentially on $d$. An exponential growth at large distance means that the iteration is unstable, an exponential decay means that it is stable. An alternative to this study is to study the SP on a single large sample and see whether it converges.

- The "instability of the second kind" of [28] amounts to a study of the proliferation of "bugs". Indeed, suppose that the input warning along link 1, which was equal to $\beta_1$ in (19), is turned to another value $\beta_0$. This is a finite change, a "bug", but we suppose that it happens with a small probability $p_{\beta_1 \to \beta_0}^1$. In a linear response, the probability $p_{\gamma \to \delta}^f$ that this will induce a bug $\gamma \to \delta$ in the output survey is:

$$p_{\gamma\to\delta}^\ell = \frac{1}{Z} \sum_{\substack{(\beta_1,\beta_2,...,\beta_n)\to\gamma \\ (\beta_0,\beta_2,...,\beta_n)\to\delta}} p_{\beta_1\to\beta_0}^1 ....p_{\beta_n}^n \ \chi(\beta_0,\beta_2...,\beta_n) \ , \tag{22}$$

which defines the matrix:

$$V_{\gamma\to\delta,\beta_1\to\beta_0} \equiv \frac{\partial p_{\gamma\to\delta}^\ell}{\partial p_{\beta_1\to\beta_0}^1} \ . \tag{23}$$

In a general situation where the warnings can have $q$ independent states, this is a square matrix of dimension $q(q-1)$. In K-SAT it is simply a $2 \times 2$ matrix. The instability to bug proliferation is determined from a product of

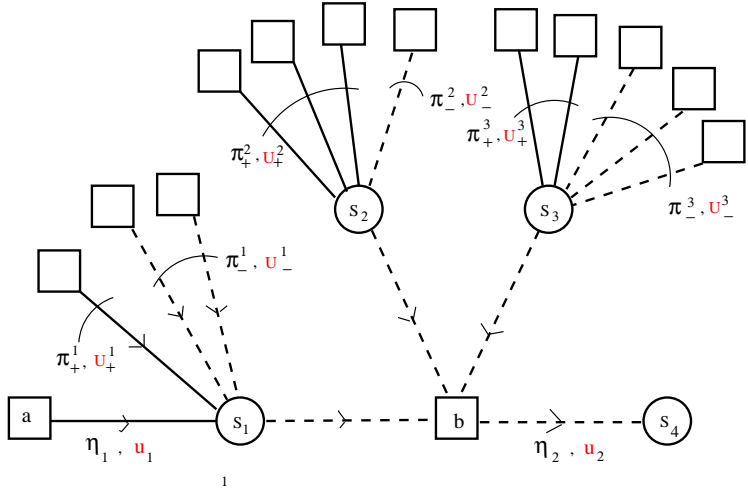

Fig. 2. Notations used in the stability analysis. The dependency of the output cavity-bias-survey (CBS) $\eta_2$ on the input CBS $\eta_1$ allows to determine the iteration stability. The propagation of a change of the input warning $u_1$ to a change in the output warning $u_2$ allows to determine the stability to bug proliferation.

$d$ such matrices

$$(K\alpha(K-1))^d Tr\langle[V^1...V^d]\rangle \equiv \mu_d . \qquad (24)$$

Note that now there is no need to average the square of the perturbation since the perturbation $p_{\beta_1 \to \beta_0}$ is always positive [28,31]. The bugs proliferate if $\mu_d$ grows exponentially with $d$, they remain localized if it decreases exponentially.

4.2 *Iteration stability*

Let us apply this general formalism to K-SAT, starting with the iteration instability. In the survey propagation we need to single out the contribution of a single survey, $\eta_1$. We follow the notations of fig.2 and write the survey propagation(2) and the Jacobian (which in this case is a real number) as:

$$\eta_2 = \frac{\left[1-(1-\eta_1)\pi_+^1\right]\pi_-^1}{(1-\eta_1)\pi_+^1 + \pi_-^1 - (1-\eta_1)\pi_+^1\pi_-^1} \prod_{i=2}^{K-1} \frac{\pi_-^i(1-\pi_+^i)}{\pi_+^i + \pi_-^i - \pi_+^i\pi_-^i} \qquad (25)$$

$$T = \frac{\partial \eta_2}{\partial \eta_1} = \frac{\pi_+^1 \pi_-^1}{[(1-\eta_1)\pi_+^1 + \pi_-^1 - (1-\eta_1)\pi_+^1\pi_-^1]^2} \prod_{i=2}^{K-1} \frac{\pi_-^i(1-\pi_+^i)}{\pi_+^i + \pi_-^i - \pi_+^i\pi_-^i} \qquad (26)$$

With respect to the general SP equation (2), we have singled out explicitly in the input the CBS $\eta_1$ from clause $a$ to variable $S_1$. Therefore the definition of variable $\pi_+^1$ differs from the general one (1) in that it does not include the factor $1-\eta_1$.

Notice that this analysis has been carried out for the special choice of clause of fig. 2, where variable $S_1$ appears negated in clause $b$ and unnegated in clause $a$ (we say that the links $S_1 \leftrightarrow b$ and $S_1 \leftrightarrow a$ are of opposite nature). In the case in which one changes for instance the edge $S_1 \to b$ to a full line, meaning that $S_1$ appears unnegated in $b$ as well as in $a$, the roles of $\pi_+^1$ and $\pi_-^1$ are exchanged, which also changes the Jacobian. One gets in this case:

$$\eta_2 = \frac{\left[1 - \pi_-^1\right](1-\eta_1)\pi_+^1}{(1-\eta_1)\pi_+^1 + \pi_-^1 - (1-\eta_1)\pi_+^1\pi_-^1} \prod_{i=2}^{K-1} \frac{\pi_-^i(1-\pi_+^i)}{\pi_+^i + \pi_-^i - \pi_+^i\pi_-^i} \tag{27}$$

$$T = \frac{\partial \eta_2}{\partial \eta_1} = -\frac{\pi_+^1 \pi_-^1 \left[1 - \pi_-^1\right]}{[(1-\eta_1)\pi_+^1 + \pi_-^1 - (1-\eta_1)\pi_+^1\pi_-^1]^2} \prod_{i=2}^{K-1} \frac{\pi_-^i(1-\pi_+^i)}{\pi_+^i + \pi_-^i - \pi_+^i\pi_-^i} \tag{28}$$

The statistical analysis of the stability proceeds as follows. Suppose one knows the pdf of the $\eta$ and $\pi$ variables, or equivalently, performing the change of variables (6), one knows $S(\phi)$ and $B(x)$, the solutions of the coupled equations (7,9). If the links $S_1 \to b$ and $a \to S_1$ are of opposite nature, the iteration equation (25) and the Jacobian (26) can be expressed as:

$$e^{-\phi_2} = 1 - \frac{e^{x_+^1 + \phi_1} - 1}{e^{x_-^1} + e^{\phi_1 + x_+^1} - 1} \prod_{i=2}^{K-1} \frac{e^{x_+^i} - 1}{e^{x_+^i} + e^{x_-^i} - 1}$$

$$T = \frac{e^{x_+^1 + x_-^1 + 2\phi_1}}{\left[e^{x_-^1} + e^{\phi_1 + x_+^1} - 1\right]^2} \prod_{i=2}^{K-1} \frac{e^{x_+^i} - 1}{e^{x_+^i} + e^{x_-^i} - 1} \tag{29}$$

If they are of the same nature, the iteration equation (27) and the Jacobian (28) can be expressed as:

$$e^{-\phi_2} = 1 - \frac{e^{x_-^1} - 1}{e^{x_-^1} + e^{\phi_1 + x_+^1} - 1} \prod_{i=2}^{K-1} \frac{e^{x_+^i} - 1}{e^{x_+^i} + e^{x_-^i} - 1}$$

$$T = \frac{e^{x_+^1 + 2\phi_1}(e^{x_-^1} - 1)}{\left[e^{x_-^1} + e^{\phi_1 + x_+^1} - 1\right]^2} \prod_{i=2}^{K-1} \frac{e^{x_+^i} - 1}{e^{x_+^i} + e^{x_-^i} - 1} \tag{30}$$

Starting from a variable $\phi_1$, chosen randomly from the distribution $S(\phi)$, and from $T_0 = 1$, using $2K - 2$ variables $x_\pm^i$ chosen randomly and independently from the distribution $A(x)$, one generates, using either (29) (with probability 1/2), or (30) (with probability 1/2), a variable $\phi_2$ and the Jacobian $T_1 = T$. This process is then iterated: using $\phi_2$ as input, and $2K - 2$ new independent random numbers $x_\pm^i$, one generates through (29) or (30) a new variable $\phi_3$ and the new Jacobian $T_2$. This process is iterated $d$ times, and one computes the total Jacobian at distance $d$, $\mathcal{T}_d = T_1 T_2....T_d$. According to the general iteration stability rule (21), one needs first to average $\mathcal{T}_d^2$ over many realizations of the random variables used in its computation, and then study the limit

$$\lim_{d \to \infty} \frac{1}{d} \ln \left( \langle \mathcal{T}_d^2 \rangle \right) + \ln(K(K-1)\alpha) , \qquad (31)$$

If this limit is negative, the system is stable; if it is positive, the system is unstable.

### 4.3 Stability to bug proliferation

We use the general formalism of (22,23) and apply it to the K-SAT problem. The stability matrix $V$ defined in (23) is a $2 \times 2$ matrix.

Let us first compute the matrix element $V_{0\to 1, 0 \to 1}$. This corresponds to introducing the bug: $u_1 = 0 \to u_1 = 1$ and computing the probability that it propagates to a bug $u_2 = 0 \to u_2 = 1$. We follow the notations of fig.2: $u_1$ is the warning going along the left link, the probability that this warning is present ($u_1 = 1$) is equal to $\eta_1$, the probability that it is absent ($u_1 = 0$) is $1 - \eta_1$. The variable $U_+^1$ is the sum of warnings arriving on variable $S_1$, indicating that it should take the value TRUE (which violates clause $b$), from function nodes distinct from $a$. The other $U$ variables are defined similarly, as the sum of warnings arriving along a selected subset of bonds indicated on the figure. In order to study $V_{0\to 1, 0 \to 1}$, the general formula (22) says that one should sum over all configurations of warnings $\{U_\pm^i\}$ such that:

- If $u_1 = 0$, the set of warnings $\{U_\pm^i\}$ automatically gives $u_2 = 0$.
- If $u_1 = 1$, the set of warnings $\{U_\pm^i\}$ automatically gives $u_2 = 1$, *and there is no contradiction* in the messages.

In order to fix the corresponding possible values of $\{U_\pm^i\}$, one can proceed as follows:

(1) In the case $u_1 = 1$, we need to send a warning $u_2 = 1$. This implies that all the variables $S_2, S_3, ..., S_{K-1}$ receive non contradictory messages assigning them to values which violate the clause $b$. Therefore $\forall i \in \{2, ..., K-1\}$: $U_+^i \geq 1$ and $U_-^i = 0$. This is the only constraint applying on these warnings, and the total probability of these warnings is $\prod_{i=2}^{K-1} \left[ \pi_-^i (1 - \pi_+^i) \right]$.
(2) The total cavity-field seen by variable $S_1$ is $U_+^1 + u_1 - U_-^1$. It should be $\leq 0$ when $u_1 = 0$ and $\geq 1$ when $u_1 = 1$. Therefore one needs $U_+^1 = U_-^1$. Furthermore there should be no contradiction when $u_1 = 1$, this implies that $U_-^1 = 0$. So the only possibility left for $U_\pm^1$ is $U_+^1 = U_-^1 = 0$. The total probability of such configurations is $\pi_-^1 \pi_+^1$.

Using (22), we thus find that the matrix element $V_{0\to 1,0\to 1}$ is:

$$V_{0\to 1,0\to 1} \equiv \tilde{v} = \frac{\pi_+^1 \pi_-^1}{(1-\eta_1)\pi_+^1 + \pi_-^1 - (1-\eta_1)\pi_+^1\pi_-^1} \prod_{i=2}^{K-1} \frac{\pi_-^i(1-\pi_+^i)}{\pi_+^i + \pi_-^i - \pi_+^i\pi_-^i} \ . \tag{32}$$

Note that here we have computed this matrix element in the case described in fig. 2. One should also study other cases in which the variables appear in the clauses with different patterns of negations, which amounts to changing some dashed lines into full lines in fig.2 and vice versa. If for instance variable $S_2$ appears in clause $b$ with a full line, the roles of $U_+^2$ and $U_-^2$ are exchanged. This is irrelevant in the following statistical analysis since $\pi_+^2$ and $\pi_-^2$ have the same distribution. It turns out that there is only one relevant possible change which alters the previous result: If the link between $S_1$ and $b$, and the one between $a$ and $S_1$ are of the same nature (for instance if $S_1$ appears negated both in $a$ and in $b$), then there is no way in which the bug $u_1 = 0 \to u_1 = 1$ can propagates to a bug $u_2 = 0 \to u_2 = 1$: the matrix element is $V_{0\to 1,0\to 1} = 0$.

We now compute the matrix element $V_{1\to 0,1\to 0}$. The inspection is very similar to the above one and it turns out that the configurations of $\{U_\pm^i\}$ which contribute to this matrix element are exactly the same as the one contributing to $V_{0\to 1,0\to 1}$. Therefore $V_{0\to 1,0\to 1} = V_{1\to 0,1\to 0}$.

We now compute the matrix element $V_{1\to 0,0\to 1}$. One should sum over all configurations of warnings $\{U_\pm^i\}$ such that:

- If $u_1 = 0$, the set of warnings $\{U_\pm^i\}$ automatically gives $u_2 = 1$.
- If $u_1 = 1$, the set of warnings $\{U_\pm^i\}$ automatically gives $u_2 = 0$, *and there is no contradiction* in the messages.

The constraints on $\{U_\pm^i\}$ are:

(1) If the link between $S_1$ and $b$, and the one between $a$ and $S_1$ are of opposite nature (as in fig. 1), increasing $u_1$ only increases the polarization of $S_1$ in the direction which violates clause $b$, and therefore the bug propagation is impossible: $V_{1\to 0,0\to 1} = 0$.
(2) If the link between $S_1$ and $b$, and the one between $a$ and $S_1$ are of the same nature (e.g. if the edge between $S_1$ and $b$ in fig. 1 becomes a full line): when $u_1 = 0$ we need $U_+^1 - U_-^1 \leq -1$ so that $S_1$ is polarized in the direction violating clause $b$; when $u_1 = 1$ we need $U_+^1 + u_1 - U_-^1 \geq 0$, and the constraint of no-contradiction imposes $U_-^1 = 0$. This contradicts the $U_+^1 - U_-^1 \leq -1$ constraint, thus there is no set of warnings propagating this bug.

In all cases: $V_{1\to 0,0\to 1} = 0$.

The matrix element $V_{0\to 1,1\to 0}$ turns out to be zero when the link $S_1 \to b$,

and the link $a \to S_1$ are of opposite natures, and nonzero whenever the link $S_1 \to b$, and the link $a \to S_1$ are of the same nature. In this last case its value $\tilde{w}$ can easily be computed using the same tools as before, but we do not need it in the stability analysis, as we now show.

To summarize, the matrix $V$ takes two values, depending on the relative nature of the links $S_1 \to b$, and $a \to S_1$:

- If these two links are of opposite nature, then

$$V = \begin{pmatrix} \tilde{v} & 0 \\ 0 & \tilde{v} \end{pmatrix} \tag{33}$$

- If these two links have the same nature, then

$$V = \begin{pmatrix} 0 & 0 \\ \tilde{w} & 0 \end{pmatrix} \tag{34}$$

The numbers $\tilde{v}, \tilde{w}$ are random numbers with distributions which can be deduced from the known distributions of the $\eta$ and $\pi$ variables.

The system is stable to bug proliferation whenever $\mu_d$ defined in (24) decreases at large $d$. We notice that the product of $d$ matrices appearing in (24) gives a null matrix whenever the second type of matrix (the off-diagonal one) appears at least twice in the product. Therefore one can forget about the off-diagonal matrix. Using (32) one generates the sequence of numbers $\tilde{v}^1...\tilde{v}^d$, from which one gets:

$$\lim_{d \to \infty} \frac{1}{d} \ln \mu_d = \ln\left(\frac{K\alpha(K-1)}{2}\right) + \lim_{d \to \infty} \frac{1}{d} \ln \langle [\tilde{v}^1...\tilde{v}^d] \rangle \;, \tag{35}$$

where the factor $1/2$ in the first ln is due to the constraint that the successive links along the stability chain must be of opposite nature, which happens with probability $1/2$. We can just deal with diagonal matrices, and $\mu_d$ is thus fixed by the average of the product of $\tilde{v}$ terms, determined in (32).

The statistical analysis of this stability to bug proliferation proceeds as follows. Performing the change of variables (6), we can use $\phi$ and $x$ variables taken randomly from the supposedly known distributions $S(\phi)$ and $A(x)$ defined in (7,9). The iteration equation (25) and the matrix element (32) can be expressed as:

$$e^{-\phi_2} = 1 - \frac{e^{x_+^1+\phi_1}-1}{e^{x_+^1+\phi_1}+e^{x_-^1}-1} \prod_{i=2}^{K-1} \frac{e^{x_+^i}-1}{e^{x_+^i}+e^{x_-^1}-1} \tag{36}$$

$$\tilde{v} = \frac{e^{\phi_1}}{\left[e^{x_+^1+\phi_1}+e^{x_-^1}-1\right]} \prod_{i=2}^{K-1} \frac{e^{x_+^i}-1}{e^{x_+^i}+e^{x_-^1}-1} \tag{37}$$

Starting from a variable $\phi_1$, chosen randomly from the distribution $S(\phi)$, and from $T_0 = 1$, using the $2K-2$ variables $x_\pm^i$ chosen randomly and independently from the distribution $B(x)$, one generates through (37) a variable $\phi_2$ and the matrix element $\tilde{v}_1 = \tilde{v}$. This process is then iterated: using $\phi_2$ as input, and $2K-2$ new random numbers $x_\pm^i$, one generates through (37) a new variable $\phi_3$ and the new matrix element $\tilde{v}_2$. This process is iterated $d$ times, and one computes $\tilde{v}_1 \tilde{v}_2 .... \tilde{v}_d$. The quantity $\langle [\tilde{v}^1 ... \tilde{v}^d] \rangle$ appearing in (35) is the average of this product over many realizations of the random variables used in its computation. If $\lim_{d\to\infty} \frac{1}{d} \ln \mu_d$ is negative the system is stable with respect to bug proliferation, if it is positive, the system is unstable.

## 5 Numerical solution of the statistical equations

### 5.1 Regularization

In order to compute the pdfs $S$ and $B$, solutions of (7,9), numerically, it is useful to first remark that they have a $\delta$ function peak at argument equal 0. The weight of this peak can be computed analytically, and substracting it allows to work with continuous random variables. We thus write:

$$\begin{aligned} S(\phi) &= t\delta(\phi) + (1-t)S_r(\phi) \\ B(x) &= \tau\delta(x) + (1-\tau)A(x) \end{aligned} \tag{38}$$

where $S_r$ and $A$ are continuous in 0. From the self-consistency equations (7,9) one gets

$$\begin{aligned} t &= 1 - (1-\tau)^{K-1} \\ \tau &= \exp\left(-\frac{K\alpha}{2}(1-t)\right) \end{aligned} \tag{39}$$

This equation has only one solution, $t = \tau = 1$ for $\alpha < \alpha_t(K)$, indicating that all fields $\phi$ and $x$ are zero. For $\alpha > \alpha_t(K)$ two non-trivial solutions appear, but only the one with the smaller value of $t$ is stable. $\alpha_t(K)$ is a decreasing function of $K$ and $\alpha_t(3) \simeq 1.6$. For large $K$ and large $\alpha$, the relevant solution $t$ goes to zero rapidly, as $t \sim (K-1)\,\mathrm{e}^{-\frac{1}{2}\alpha K}$.

The self-consistent set of equations for the pdfs $S_r$ and $A$ and the mixture $B$

reads

$$S_r(\phi) = \int \prod_{i=1}^{K-1} \left[dx_i dy_i\, A(x_i)B(y_i)\right] \delta\left(\phi + \ln\left[1 - \prod_{i=1}^{K-1} \frac{e^{x_i}-1}{e^{x_i}+e^{y_i}-1}\right]\right). \quad (40)$$

$$A(x) = \frac{1}{e^\gamma - 1} \sum_{k=1}^{\infty} \frac{\gamma^k}{k!} \int \prod_{j=1}^{k} \left[d\phi_j\, S_r(\phi_j)\right] \delta(x - \sum_j \phi_j) \quad (41)$$

$$B(y) = e^{-\gamma} \sum_{k=0}^{\infty} \frac{\gamma^k}{k!} \int \prod_{j=1}^{k} \left[d\phi_j\, S_r(\phi_j)\right] \delta(y - \sum_j \phi_j) \quad (42)$$

where $\gamma = \frac{K\alpha}{2}(1-t)$ is related to the solution $t$ of (39). Once $B(x)$ is known the complexity $\Sigma$ can be calculated according to (18)-(13).

## 5.2 Population Dynamics

$S_r(\phi)$, $A(x)$ and $B(y)$ are solutions of the set of integral equations (40), (41) and (42). The numerical solution of these equation proceeds in an iterative manner. Based on a first approximation of $S_r$, approximations of $A$ and $B$ are calculated using Eqs. (41) and (42). These functions are plugged into Eq. (40) to get new approximation of $S_r(\phi)$. This process is iterated until $S_r$, $A$ and $B$ attain their fixpoints. Since $S_r(\phi)$, $A(x)$ and $B(y)$ are probability densities, they can be approximated numerically by a large set of independent variables drawn from the respective density. This way the iteration of Eqs. (40), (41) and (42) becomes an iterative update of a "population" of $N$ variables $\phi_1, \ldots, \phi_N$. For given $N$, $\alpha$ and $K$ the population dynamics algorithm reads:

(1) Compute $t$ as the solution of Eq. (39). Set $\gamma = \frac{K\alpha}{2}(1-t)$.
(2) Initialize the $\phi_j$ as positive i.i.d. random variables with an exponential distribution of mean $2^{1-K}$.
(3) For $j = 1$ to $K - 1$:
   (a) Generate a random integer $k \geq 1$ with distribution $\propto \gamma^k/k!$
   (b) Pick $k$ integers $i_1, \ldots, i_k$ at random from $\{1, \ldots, N\}$.
   (c) Calculate the sum $x_j = \phi_{i_1} + \cdots + \phi_{i_k}$. $x_j$ is a random variable with distribution $A(x)$.
(4) For $j = 1$ to $K - 1$:
   (a) Generate Poisson distributed random integer $k$ with mean $\gamma$.
   (b) If $k = 0$, set $y_j = 0$.
   (c) If $k > 0$, pick $k$ integers $i_1, \ldots, i_k$ at random from $\{1, \ldots, N\}$ and set $y_j = \phi_{i_1} + \cdots + \phi_{i_k}$. $y_j$ is a random variable with distribution $B(y)$.
(5) Calculate $z = \prod_{j=1}^{K-1} \left(1 + \frac{e^{y_j}}{e^{x_j}-1}\right)$.
(6) Replace a randomly chosen variable $\phi_\ell$ in the population by the new value $\phi_0 = \ln(1 + 1/(z-1))$.

Steps 3 to 6 have to be repeated until the population of variables $\phi_\ell$ is distributed according to the stationary distribution $S_r(\phi)$. As a criterion of convergence we monitor the first (empirical) moments of the fields $\phi$: if these change little after $N$ variables $\phi$ have been updated, we might assume convergence.

After convergence has been reached (transient iterations), steps 3 to 6 are iterated $TN$ times for some large $T$. The $TN(K-1)$ random variables $y_j$ calculated in step 4 are used to estimate the complexity $\Sigma$ by approximating the integrals in (18) by sums.

For larger values of $K$ the proper initialization in step 2 is essential. Note that in step 5 quantities $\exp(x)$ with $x$ of order $\frac{K\alpha}{2}\mathbb{E}(\Phi)$ have to be calculated. Now $\alpha$ scales like $2^K$, so $\mathbb{E}(\Phi)$ should better scale like $2^{-K}$ to avoid numerical overflow. The value $\mathbb{E}(\Phi) = 2^{1-K}$ chosen in the initialization step is small enough to prevent overflow and large enough to stay away from the trivial solution (all $\Phi = 0$). In addition, initializing with $\mathbb{E}(\Phi) = 2^{1-K}$ agrees with the asymptotic behavior of $S(\Phi)$, see Sec. 6.1.

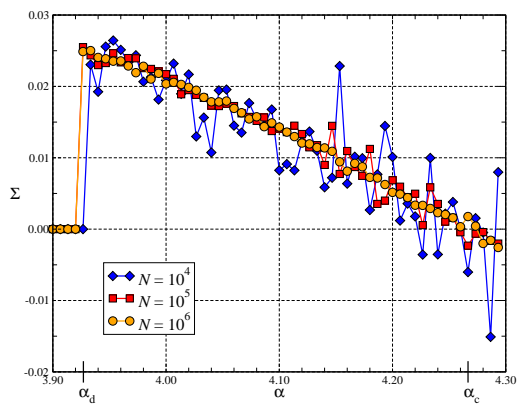

Fig. 3. Complexity $\Sigma(\alpha)$ for $K = 3$ obtained from the population dynamics algorithm with $T = 1000$. The number of transient iterations is $100N$ in all cases.

### 5.3 Threshold values

For $\alpha < \alpha_d$ the whole population of fields collapses to the value 0. We identify $\alpha_d$ as the first value of $\alpha$ where we find non-trivial stable pdf $B(x)$ and $S_r(\phi)$.

The satisfiability threshold $\alpha_c$ is studied using the complexity $\Sigma$. Fig. 3 shows the complexity for $K = 3$ and $3.9 < \alpha < 4.3$ obtained with the population dynamics. The overall shape of $\Sigma(\alpha)$ is clearly visible, but the fluctuations are strong, even for $N = 10^6$. The location $\alpha_d$ of the jump in the $\Sigma$ is easier to pin down accurately than the value $\alpha_c$ where $\Sigma(\alpha_c) = 0$.

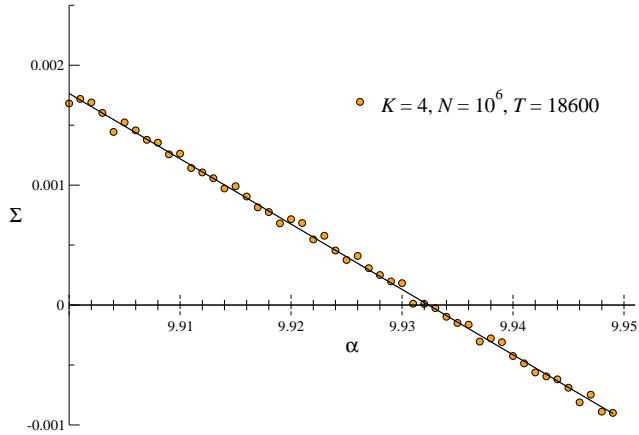

Fig. 4. Complexity $\Sigma(\alpha)$ for $K = 4$ in the vicinity of $\alpha_c$. Linear regression yields the value $\alpha_c = 9.9324$.

To get $\alpha_c$ we run the population dynamics for 50 equidistant values $\alpha$ around the estimated value of $\alpha_c$ and use linear regression to locate the root of $\Sigma(\alpha)$, see Fig. 4 for an example with $K = 4$.

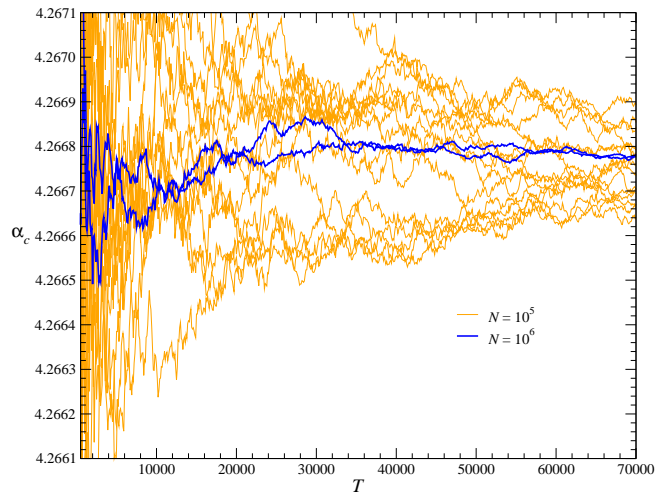

Fig. 5. Values of $\alpha_c$ as determined from various runs of the population dynamics. The convergence for $T \to \infty$ is clearly visible as well as the fluctuations induced by the finiteness of $N$. Note the resolution of the $\alpha_c$-axis.

With large values of $T$ and $N$ and repeated runs, fairly accurate estimates for $\alpha_c$ can be obtained, as well as an estimate of the error bar, see Fig. 5. Note that in a *single run* with $N = 10^6$ and $T = 70000$ more than $10^{14}$ random numbers are consumed. This makes high demands on the pseudo random number generator. We used LCG64, a linear congruential generator with period $2^{64} - 1 \approx 10^{19}$ from the TRNG-library [3]. The results have been checked with an explicit inverse generator from the same library.

| $K$ | $\alpha_d$ | $\alpha_d^{(0)}$ | $\alpha_s$ | $\alpha_c$ | $\alpha_c^{(1)}$ | $\alpha_c^{(2)}$ | $\alpha_c^{(7)}$ |
|---|---|---|---|---|---|---|---|
| 3 | **3.927** ± 0.004 | 3.923 | **4.15** | **4.267** | 4.699 | 4.546 | 4.307 |
| 4 | **8.297** ± 0.008 | 8.303 | **9.08** | **9.931** | 10.244 | 10.104 | 9.938 |
| 5 | **16.12** ± 0.02 | 16.117 | **17.8** | **21.117** | 21.334 | 21.223 | 21.118 |
| 6 | **30.50** ± 0.03 | 30.479 | **33.6** | **43.37** | 43.515 | 43.434 | 43.372 |
| 7 | **57.22** ± 0.06 | 57.186 | **62.5** | **87.79** | 87.876 | 87.821 | 87.785 |
| 8 | **107.24** ± 0.08 | 107.191 | | | 176.599 | 176.563 | 176.543 |
| 9 | **201.35** ± 0.1 | 201.276 | | | 354.045 | 354.022 | 354.010 |
| 10 | **379.10** ± 0.1 | 379.004 | | | 708.936 | 708.922 | 708.915 |

Table 1
Threshold values for random $K$-SAT. Bold numbers are the results of the population dynamics algorithm. $\alpha_d^{(0)}$ is the value predicted by the first moment expansion of the cavity equations (sec. 6.3), $\alpha_c^{(r)}$ is the result of a series expansion in $\varepsilon = 2^{-K}$ of the cavity equations up to order $r$ (secs. 6.2 and A). Note that all reported values $\alpha_c(K)$ fall between the best rigorously known upper and lower bounds.

Table 1 shows the results. Since $\alpha_c$ for $K = 3$ is the most "prominent" threshold we spent a bit more CPU power to increase its accuracy. Currently our best estimate is

$$\alpha_c(3) = 4.26675 \pm 0.00015 \tag{43}$$

The errorbars in table 1 and in Eq. (43) are given by $\pm 2\sigma$, where $\sigma$ is the empirical standard deviation as measured by different runs of the population dynamics algorithm (see Fig. 5) with fixed $N$ and $T$. The quoted values of $\alpha_c$ are the empirical averages over different runs. The simulations show that the averages are not very sensitive to the value of $N$. The errorbars on the other hand get smaller with increasing $N$.

### 5.4 Stability

Using the population of $\phi$ and $x$ variables obtained in the population dynamics, one can check the stability of this 1-RSB solution. In all cases we find that for $\alpha \leq \alpha_c$ the solution is stable to iteration. As for the bug proliferation stability, we estimate its location by considering the value of $\mu_d$ defined in (24). In Fig.6 we plot $\ln \mu_d$ versus $d$ for the case $K = 4$, for various values of $\alpha$. The behaviour is well approximated by a linear function. Using a linear regression, we estimate the slope and plot it as function of $\alpha$, as shown in Fig. 7. In this way we estimate the limit of stability of the 1-RSB solution to: $\alpha_s(4) \simeq 9.08$. The values of $\alpha_s(K)$ for $K = 3, \ldots, 7$ are shown in Tab. 1.

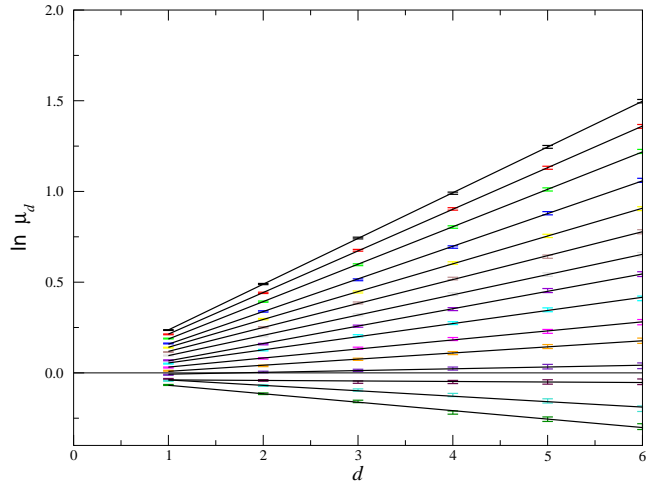

Fig. 6. Stability of the 1RSB solution to the $K=4$ random satisfiability problem. The stability parameter $\ln\mu_d$ of (24) is plotted versus the distance $d$ for various values of $\alpha$ equally spaced between $\alpha = 8.50, 8.55, ..., 9.20$ ( from top to bottom). The points with error bars are the results of the numerical evaluation of $\mu_d$ using the population dynamics algorithm, the lines are the best linear fits.

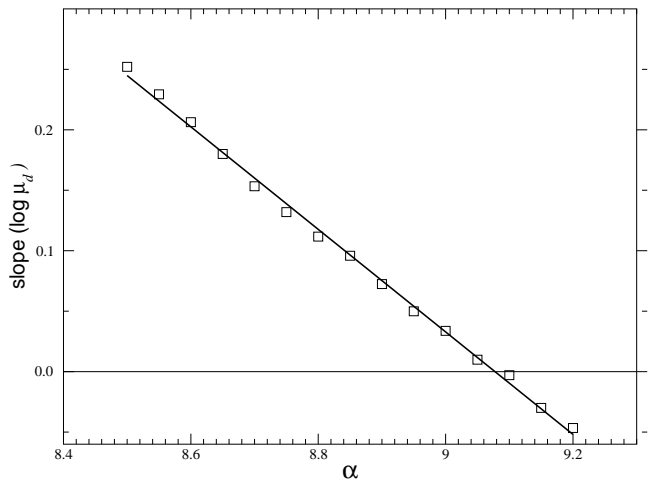

Fig. 7. The slopes of the linear fits to $\ln\mu_d$ versus $d$, obtained in fig.6 for the $K=4$ random satisfiability problem, are plotted versus $\alpha$. Values of $\alpha$ such that the slope is negative are such that the 1RSB solution is stable.

## 6  Large $K$ analysis

### 6.1  Introduction

In the large $K$ limit the random $K$-sat problem simplifies and one can get more precise analytic results. One knows from [8,2] that, at first order in the expansion in powers of $2^{-K}$, the rigorous bounds scale as

$$2^{-K}\alpha_{\mathrm{LB}}(K) \simeq \ln(2) - 2^{-K}\left(\frac{K+1}{2}\ln(2) + 1 + o(1)\right) + \mathcal{O}(2^{-2K}) \qquad (44)$$

$$2^{-K}\alpha_{\mathrm{UB}}(K) \simeq \ln(2) - 2^{-K}\frac{1+\ln(2)}{2} + \mathcal{O}(2^{-2K}) \qquad (45)$$

where $o(1)$ is a term which goes to 0 at large $K$. It is interesting also to note that the best lower bounds obtained by studying simple algorithms give a scaling $2^{-K}\alpha_{\mathrm{LB}}(K) \simeq c/K$ where $c$ is a constant [12].

In this section we shall analyse the cavity equations (40-42) in the large $K$ limit using an expansion in the small parameter $\varepsilon = 2^{-K}$, and defining the rescaled parameter

$$\hat{\alpha} = 2^{-K}\alpha. \qquad (46)$$

We shall compute the large $K$ behavior of the satisfiability threshold $\alpha_c$, the dynamic threshold $\alpha_d$ and the stability threshold $\alpha_s$. The expansion will be in powers of $\varepsilon$, although one must be careful that at each order powers of $K$ also appear.

We need to find an expansion of the pdfs $S_r$, $A$ and $B$ which satisfy the self-consistent equations (40-42). We first note that at large $K$ and large $\alpha$, the solution $t$ of (39) goes to zero as $t \sim (K-1)\exp(-K\alpha/2)$, more rapidly than any power of $\varepsilon$. We can thus take $t = 0$ in our large $K$ analysis. Then $\gamma = (K/2)2^K\hat{\alpha} \sim 1/\varepsilon$ is large.

According to Eq.(42) $y$ is a sum of $k$ fields $\phi$, where $k$ is Poisson distributed with mean $\gamma$. Eq.(41) shows that the distribution $A(x)$ is very close to the distribution $B(x)$, up to corrections of order $\exp(-\gamma)$ which again can be neglected in an expansion in powers of $\varepsilon$. Hence we expect that for $K \gg 1$ $A(x) \simeq B(x)$ become Gaussians concentrated around $\gamma M_1$, where $M_1 = \mathbb{E}(\phi)$. In the same regime, the main contributions to the integral in (40) come from the constant term $2^{1-K}$, hence we expect $S_r(2^{K-1}\phi)$ to be concentrated around 1: then $M_1 \simeq 2^{1-K}$ and $\gamma M_1 \simeq K\hat{\alpha}$. This scenario is confirmed by the analysis of the functions $A$, $B$, and $S_r$ obtained from the population dynamics. We will use it as the starting point for an asymptotic expansion for $\alpha_c(K)$ and for the large $K$ asymptotics of $\alpha_d$ and $\alpha_s$.

6.2 *Asymptotic Expansion for $\alpha_c$*

Our goal is to express $\hat{\alpha}_c(K) = \alpha_c(K)\, 2^{-K}$ as a power series in the variable $\varepsilon = 2^{-K}$, with coefficients which may contain powers of $K$ or of $\ln K$. We know from the bounds (45) that $\hat{\alpha}_c = \ln(2) + \mathcal{O}(\varepsilon)$. To get the next terms of the series we need to expand all moments of $A \simeq B$ and $S_r$ in powers of $\varepsilon$.

Let $M_n = \mathbb{E}(\phi^n) = \int d\phi\, S_r(\phi)\phi^n$ denote the $n$-th moment of $S_r$. Concerning the distribution $B(y)$, it is useful to introduce three types of moments. The $n$-th moment of $B$ is $\mathbb{E}(y^n) = \int dy\, B(y)y^n$. The $n$-th cumulant of $B$, $\hat{b}_n$, is defined as usual by the identity between generating functions:

$$\exp\left[\sum_{n=1}^{\infty} \frac{\lambda^n}{n!}\hat{b}_n\right] = \sum_{n=0}^{\infty} \frac{\lambda^n}{n!}\mathbb{E}(y^n) \ . \tag{47}$$

The $n$-th central moment of $B$, $b_n$, is defined by $b_n = \mathbb{E}((y - \mathbb{E}(y))^n)$. Eq. (42) shows that

$$\hat{b}_n = \gamma M_n \quad n \geq 1 \ . \tag{48}$$

As motivated in the introduction of this section we will assume that $M_n = \mathcal{O}(\varepsilon^n)$. From

$$\gamma = \frac{\hat{\alpha}K}{2\varepsilon}\left(1 - \mathcal{O}(\varepsilon^{2^{K-1}})\right), \tag{49}$$

we get the scaling of the cumulants $\hat{b}_n$ and the central moments $b_n$ as

$$\begin{aligned} \hat{b}_n &= \mathcal{O}(\varepsilon^{n-1}) & n \geq 1 \\ b_n &= \mathcal{O}(\varepsilon^{\lceil n/2 \rceil}) & n \geq 2, \end{aligned} \tag{50}$$

where $\lceil x \rceil$ denotes the smallest integer $\geq x$. These scalings are the base of our asymptotic expansion. To calculate the moments $b_n$ self-consistently we rewrite $S_r(\phi)$ as

$$S_r(\phi) = \int \prod_{i=1}^{K-1} \left[d\zeta_i\, D(\zeta_i)\right] \delta\left(\phi - \ln\left[\frac{1}{1 - \prod_{i=1}^{K-1}\zeta_i}\right]\right) \tag{51}$$

with

$$D(\zeta) = \int dx\, dy\, B(x)B(y)\delta\left(\zeta - \frac{e^x - 1}{e^x + e^y - 1}\right) \ . \tag{52}$$

The moments of $D$ are given by

$$\mathbb{E}\left(\zeta^\ell\right) = \sum_{p,q=0}^{\infty} D_{p,q}^{(\ell)} b_p b_q \tag{53}$$

where $D_{p,q}^{(\ell)}$ is the Taylor coefficient

$$D_{p,q}^{(\ell)} = \frac{1}{p!\, q!} \frac{\partial^{p+q}}{\partial^p x\, \partial^q y} \left(\frac{e^x - 1}{e^x + e^y - 1}\right)^\ell \bigg|_{x=y=\mathbb{E}(y)} \ . \tag{54}$$

As we know that $b_0 = 1$, $b_1 = 0$ and $b_{n\geq 2} = \mathcal{O}(\varepsilon^{\lceil n/2 \rceil})$, we find that

$$\mathbb{E}\left(\zeta^\ell\right) \simeq D_{0,0}^{(\ell)} + \mathcal{O}(\varepsilon) = \left(\frac{1-\delta}{2-\delta}\right)^\ell + \mathcal{O}(\varepsilon) \ , \tag{55}$$

where we have introduced $\delta = \exp(-\mathbb{E}(y))$. Because the leading contribution to $\gamma M_1$ is $\gamma M_1 \simeq K\hat{\alpha}$, and $\hat{\alpha}$ is close to $\ln 2$, we expect $\delta$ to be a small parameter of order $\varepsilon$. We thus write:

$$D_{0,0}^{(\ell)} = \left(\frac{1-\delta}{2-\delta}\right)^\ell = 2^{-\ell}\left(1 - \frac{\ell}{2}\delta\right) + \mathcal{O}(\varepsilon^2) \tag{56}$$

$$D_{2,0}^{(\ell)} + D_{0,2}^{(\ell)} = 2^{-\ell}\frac{\ell(\ell-1)}{4} + \mathcal{O}(\varepsilon) \tag{57}$$

Expanding $-\ln(1-x)$ in the integral in (51) in a Taylor series at $x = 0$ gives

$$\begin{aligned}
M_1 &= \sum_{\ell=1}^{\infty} \frac{1}{\ell} \mathbb{E}\left(\zeta^\ell\right)^{K-1} \\
&= \sum_{\ell=1}^{\infty} \frac{(2\varepsilon)^\ell}{\ell}\left(1 - \frac{\ell}{2}\delta + \frac{\ell(\ell-1)}{4}b_2 + \mathcal{O}(\varepsilon^2)\right)^{K-1} \\
&= 2\varepsilon\left(1 - \frac{K-1}{2}\delta\right) + 2\varepsilon^2 + \mathcal{O}(\varepsilon^3)
\end{aligned} \tag{58}$$

and similarly
$$M_2 = 4\varepsilon^2 + \mathcal{O}(\varepsilon^3). \tag{59}$$

Since $\hat{b}_1 = \mathbb{E}(y)$ and $\hat{b}_2 = b_2$ we can use Eq. (48) to get the desired series expansion for the moments $b_n$:

$$\mathbb{E}(y) = \hat{\alpha}K - \hat{\alpha}\frac{K(K-1)}{2}\varepsilon + \hat{\alpha}K\varepsilon + \mathcal{O}(\varepsilon^2) \tag{60}$$

$$b_2 = 2\hat{\alpha}K\varepsilon + \mathcal{O}(\varepsilon^2) \tag{61}$$

Now we are ready to expand the complexity into a series in $\varepsilon$. We use the representation of the complexity in terms of integrals over $B$, given in (18)–(13). Again we expand the logarithms in both integrals in a Taylor series around $\mathbb{E}(y)$ to get

$$\Sigma_0 = \ln(2) + \hat{\alpha}K - \hat{\alpha}\frac{K^2}{2}\varepsilon + 2\hat{\alpha}K\varepsilon - \frac{\delta}{2} + \mathcal{O}(\varepsilon^2) \tag{62}$$

and

$$K I_{K-1} - (K-1) I_K = -(K+1)\varepsilon - \frac{3K+1}{2}\varepsilon^2 + \frac{K(K-2)}{2}\varepsilon^2 + \mathcal{O}(\varepsilon^3). \tag{63}$$

From (18) we get $\Sigma(\alpha)$ up to order $\mathcal{O}(\varepsilon^2)$, and solving for $\hat{\alpha}_c$ we finally arrive at

$$\hat{\alpha}_c = \ln(2) - \frac{\ln(2)+1}{2}\varepsilon + \mathcal{O}(\varepsilon^2). \tag{64}$$

Note that this result is identical to the best known *rigorous upper bound* for $\alpha_c$ found in [8,17].

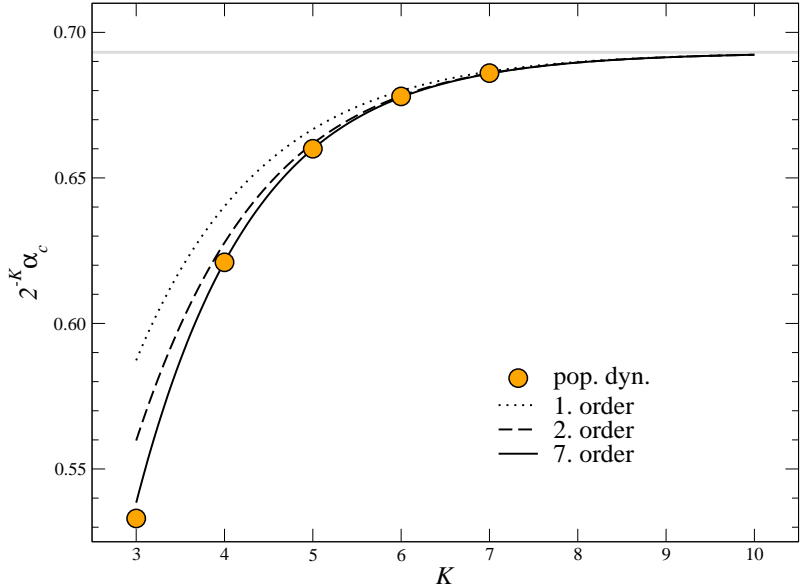

Fig. 8. Statistical transition point $\hat{\alpha}_c = \alpha_c\, 2^{-K}$: population dynamics solution of the cavity equations compared to the asymptotic expansions of first, second and seventh order in $\varepsilon = 2^{-K}$, see appendix A.

In appendix A we show how to calculate higher order terms of the asymptotic expansion for $\alpha_c$. Fig. 8 and Table 1 show that the seventh order expansion gets actually very close to the numerical values. For $K = 3$ the deviation of the seventh order asymptotic expansion from the numerical value is less than 1%, and for $K \geq 4$ this deviation is even smaller. Already at second order the cavity 1-RSB result is strictly smaller than the best known upper bound of [8] (which is itself better than the upper bound of [17]).

### 6.3  Asymptotics of $\alpha_d$

The distributions $S(\phi)$, $A(x)$ and $B(y)$ get more and more concentrated as $K$ gets larger. For an asymptotic analysis of $\alpha_d$ we will treat all pdfs as $\delta$-functions. In terms of the moment expansion this means to ignore all but the first cumulant of $B$. We shall first work out the resulting value that is obtained from such an approximation for any $K$, and then compute the large $K$ asymptotics.

If $S(\phi)$ is concentrated at $M_1$, then $B$ and $A$ are concentrated at $\mathbb{E}(y) = \gamma M_1$ and $\mathbb{E}(x) = \gamma M_1/(1 - e^{-\gamma})$. Eq. (40) gives

$$M_1 = \ln\left[1 + \frac{(e^{\mathbb{E}(x)} - 1)^{K-1}}{(e^{\mathbb{E}(x)} + e^{\mathbb{E}(y)} - 1)^{K-1} - (e^{\mathbb{E}(x)} - 1)^{K-1}}\right], \qquad (65)$$

With $z = e^{-\gamma M_1}$ this can be written as

$$z = f(z) = \left[1 - \left(\frac{1 - z^{\frac{1}{1-e^{-\gamma}}}}{1 + z^{\frac{1}{e^\gamma - 1}} - z^{\frac{1}{1-e^{-\gamma}}}}\right)^{K-1}\right]^\gamma. \tag{66}$$

The trivial $z = 1$ is always a solution of this equation, but for $\gamma > \gamma_d^{(0)}(K)$ non trivial solutions $z < 1$ appear. The critical value $\gamma_d^{(0)}(K)$ is given by the solution of

$$z = f(z) \qquad \text{and} \qquad 1 = f'(z). \tag{67}$$

These equations can be solved numerically and the resulting value $\gamma_d^{(0)}$ is easily translated into a corresponding $\alpha_d^{(0)}$ by Eq. (39). The results are the values for $\alpha_d^{(0)}$ in Table 1. The values for $\alpha_d^{(0)}$ agree perfectly with the exact values $\alpha_d$ (within the error bars of the latter), even for $K = 3$, although the non-trivial distributions $A(x)$ and $B(y)$ that appear right above $\alpha_d$ are not $\delta$-like.

To analyze $\alpha_d^0$ in the large $K$-limit we simplify $f(z)$ by ignoring the difference between $\mathbb{E}(x)$ and $\mathbb{E}(y)$ as discussed in sec. 6.2. This leads to

$$z = f(z) = \left[1 - \left(\frac{1-z}{2-z}\right)^{K-1}\right]^\gamma. \tag{68}$$

The non-trivial solution of (68) is a number $z \in (0, 1)$, hence $\frac{1-z}{2-z} < \frac{1}{2}$ and for large $K$ we can write

$$z = f(z) \simeq e^{-\gamma\left(\frac{1-z}{2-z}\right)^{K-1}} \tag{69}$$

This equation should have only the trivial solution $z = 1$ for $\gamma < \gamma_d$ and a non-trivial solution $z < 1$ for $\gamma > \gamma_d$. From the numerics we know that $zK$ is small and decreasing with increasing $K$, hence we try the ansatz

$$zK = e^{-d(K)}, \tag{70}$$

where $d$ is a slowly growing function of $K$. This ansatz allows us to write

$$\left(\frac{1-z}{2-z}\right)^{K-1} \simeq 2^{1-K} e^{-\frac{e^{-d}}{2}} \tag{71}$$

and together with (69) we get

$$\gamma = 2^{K-1} \underbrace{e^{\frac{e^{-d}}{2}} (\ln K + d)}_{=: g(d)}. \tag{72}$$

The critical value $\alpha_d = 2\gamma_d/K$ is determined by the minimum of $g(d)$,

$$\alpha_d = \frac{2^K}{K} \left(\ln K + d^\star\right) e^{\frac{e^{-d^\star}}{2}} \tag{73}$$

where $d^\star$ denotes the solution of $g'(d) = 0$, which is the solution of

$$\exp(d^\star) = \frac{1}{2}(\ln K + d^\star) \tag{74}$$

$$d^\star = \ln\left(\frac{1}{2}\ln K + \frac{1}{2}d^\star\right). \tag{75}$$

Note that $d^\star = \ln\left(\frac{1}{2}\ln K\right)\left[1 + \mathcal{O}(\frac{\ln \ln K}{\ln K})\right]$ in agreement with our ansatz of a slowly growing $d(K)$, Eq. (70). Fig. 9 verifies that (73) indeed gives the correct asymptotic expansion of $\alpha_d^0$.

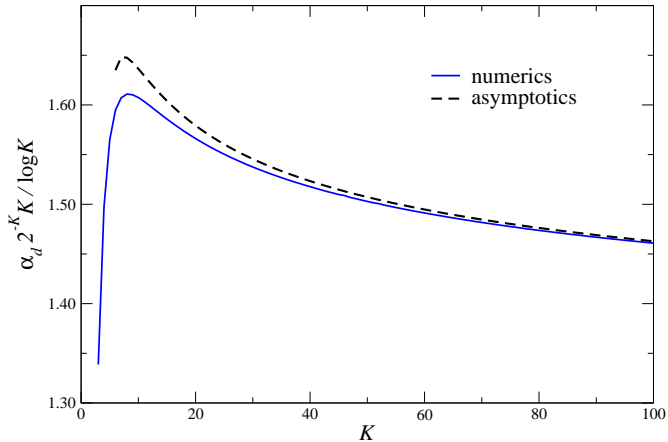

Fig. 9. Dynamical threshold $\alpha_d$: asymptotical expansion (73) compared to the numerical solution of (68).

### 6.4 Stability Analysis at Large K

Since we are interested in studying the instability of type II in the region $\alpha_d < \alpha$, let us parametrize $\alpha$ as

$$\alpha = \frac{2^K}{K}(\ln K + d)\, e^{\frac{e^{-d}}{2}}, \tag{76}$$

with $d \geq d^\star(K)$.

In the large $K$ limit the stability analysis is greatly simplified with respect to the general case discussed in Sec. 4. The iteration equation (36) and the matrix element eq. (37) read:

$$\tilde{v}_1 = \frac{e^{-(x_+^1 + x_-^1)}}{e^{-x_+^1 - \phi_1} + e^{-x_-^1} - e^{-(x_+^1 + x_-^1 + \phi_1)}} e^{-\xi_1} \tag{77}$$

and
$$e^{-\phi_2} = 1 - \frac{[1 - e^{-(x_+^1 + \phi_1)}]e^{-x_-^1}}{e^{-x_+^1 - \phi_1} + e^{-x_-^1} - e^{-(x_+^1 + x_-^1 + \phi_1)}} e^{-\xi_1} \tag{78}$$

where
$$\xi_1 = \sum_{i=2}^{K} \zeta_i \ , \quad \text{and} \quad \zeta_i = -\ln\left(\frac{e^{u_i} - 1}{e^{u_i} + e^{v_i} - 1}\right) . \tag{79}$$

At large $K$, the variables $u_i$ and $v_i$ are random variables distributed according to the law $A$.

In order to understand the simplification taking place at large $K$, it is useful to carry one more step of the iteration: To compute the stability at distance 2, we need to compute

$$\tilde{v}_2 \frac{e^{-(x_+^2 + x_-^2)}}{e^{-x_+^2 - \phi_2} + e^{-x_-^2} - e^{-(x_+^2 + x_-^2 + \phi_2)}} e^{-\xi_2} \tag{80}$$

and
$$e^{-\phi_3} = 1 - \frac{[1 - e^{-(x_+^2 + \phi_2)}]e^{-x_-^2}}{e^{-x_+^2 - \phi_2} + e^{-x_-^2} - e^{-(x_+^2 + x_-^2 + \phi_2)}} e^{-\xi_2}, \tag{81}$$

where $\xi_2$ is iid with $\xi_1$.

In general $\tilde{v}_1$ and $\tilde{v}_2$ are correlated because $\tilde{v}_2$ depends on $\phi_2$, which itself depends on the same random variables as $\tilde{v}_1$. At large $K$, we notice that a sum like $x_+^2 + \phi_2$ simplifies, because $<x> = \frac{K\alpha}{2} M_1 = O(M_1 2^K)$ and $\phi = O(M_1)$: neglecting corrections of order $\varepsilon$, we can write at leading order $x_+^2 + \phi_2 = x_+^2$. Performing this simplification every time that a sum of the type $x + \phi$ appears, we deduce that the correlations between $\tilde{v}_1$ and $\tilde{v}_2$ disappear at large $K$. The stability condition (35) simplifies to

$$\lim_{d \to \infty} \frac{1}{d} \ln \mu_d = \ln\left(\frac{K\alpha(K-1)}{2}\right) + \ln \mathbb{E}(\tilde{v}_1) . \tag{82}$$

To leading order we approximate $B(x)$ as a $\delta$-function at $x = (K\alpha/2)M_1$, so that $z = \exp(-x) = \exp(-(K\alpha/2)M_1)$ satisfies equation (68). We get

$$\mathbb{E}(\tilde{v}_1) = \frac{z}{2-z}\left(\frac{1-z}{2-z}\right)^{K-2} \tag{83}$$

Using the asymptotic form (76), together with $z = \exp(-d)/K$ and (71), in eq. (82) we find
$$e^{-d}(\ln(K) + d) = 1 . \tag{84}$$

The solution $d_s(K)$ of this equation, when plugged into the definition (76) of $\alpha$, gives the stability threshold $\alpha_s$, such that the 1-RSB cavity solution is stable at $\alpha > \alpha_s$. We notice that (84) is similar to eq. (74) giving $d^\star(K)$. In fact a simple change of variables gives:

$$d_s(K) = d^\star(2K) + \ln 2 \ . \tag{85}$$

The asymptotic expression for the instability is thus given by

$$\alpha_s = \frac{2^K}{K}(\ln(2K) + d^\star(2K))e^{\frac{1}{4}e^{-d^\star(2K)}} \tag{86}$$

which is close to the dynamical point $\alpha_d$ of (73).

For any finite $K$, there exists a region between $\alpha_d$ and $\alpha_s$ in which the one step solution is unstable, while the solution at $\alpha_s < \alpha \le \alpha_c$ is stable. At large $K$ the unstable region is small and $\lim_{K\to\infty} \frac{\alpha_d(K)}{\alpha_s(K)} = 1$. The relative difference of $(\alpha_s - \alpha_d)/\alpha_d$ is of order $1/\ln K$.

## 7 Conclusions

In this contribution we have studied in details the 1-RSB cavity analysis of the random $K$-SAT problem for general $K$. Starting from the Survey Propagation equations we have computed the threshold values $\alpha_d(K)$ for the onset of clustering of constrained variables, $\alpha_s(K)$ for the stability of the 1-RSB cavity solution, $\alpha_c(K)$ for the SAT/UNSAT transition. For all studied values of $K$ we find $\alpha_d(K) < \alpha_s(K) < \alpha_c(K)$, which shows that the SAT/UNSAT threshold $\alpha_c(K)$ always lies in the stable region. This provides stronger support to the conjecture that the 1-RSB cavity values for $\alpha_c(K)$ are exact for any $K$. On the other hand, the neighborhood of $\alpha_d(K)$ should be studied with higher order RSB methods.

We can summarize the main conjectures obtained from the statistical physics approach as follows:

**Conjecture 1:** *The satisfiability threshold $\alpha_c(K)$ for random $K$-SAT is obtained by the following procedure:*

(1) *Given $\alpha$, find the two probability distribution functions $S(\phi)$ and $B(x)$ which satisfy the integral equations (7,9).*
(2) *Compute the complexity $\Sigma$ given in (18–13).*
(3) *$\alpha_c(K)$ is the value where $\Sigma$ vanishes.*

We have developed two different approaches to computing the threshold, a numerical study of the cavity equation for $K \in \{3,...,7\}$ on one hand and a large $K$ expansion on the other hand. The large $K$ expansion can be summarized in:

***Conjecture 2:*** *The satisfiability threshold $\alpha_c(K)$ for random K-SAT can be obtained from an asymptotic expansion in $\varepsilon = 2^{-K}$, i.e. for each non-negative integer $M$, we have that*

$$2^{-K}\alpha_c(K) = \ln(2) + \sum_{i=1}^{M} \hat{\alpha}_i \varepsilon^i + o(\varepsilon^M),  \qquad (87)$$

*where the coefficients $\hat{\alpha}_i$ are polynomials in $K$ of degree $2i - 2$. Our method provides an explicit conjecture for these coefficients (see Sect. A): the first three terms are given in Eqs. (A.21–A.23), coefficients up to $\hat{\alpha}_7$ are available upon request.*

As one can see in Fig. 8, the result of the asymptotic expansion up to 7th order nicely match the direct numerical study of Eqs. (18,40–42) for $K \in \{3, \ldots, 7\}$.

On top of having the threshold value for any $K$, this analysis is also useful since the large $K$ limit is a natural candidate for building a complete analytical matching between the cavity solution and rigorous results. To the leading order the analytical expression for the threshold value at large $K$ coincides with the best known rigorous upper bounds while higher orders give a further improvement. We hope that the type of results discussed in this paper will help in developing a complete rigorous understanding of the cavity approach in the $K$-SAT problem and of its algorithmic potentialities.

## A  Systematic Moment Expansion

The series expansion for $\alpha_c(K)$ from section 6.2 can be worked out systematically. Eqs. (41), (42), (51) and (52) define a circular mapping $S_r(\phi) \mapsto [A(x), B(y)] \mapsto D(\zeta) \mapsto S_r(\phi)$ that we want to express in terms of moments. Eq. (48) is the moment version of $S \mapsto B$. As discussed in section 6.2 we will neglect the difference between $A$ and $B$ and we set $\gamma = \alpha K/2$. In terms of moments the map $D \mapsto S_r$ reads

$$M_n = \sum_{\ell=1}^{\infty} C_{n,\ell} \mathbb{E}\left(\zeta^\ell\right)^{K-1} \qquad (A.1)$$

where the Taylor coefficients $C_{n,\ell}$ are defined by $\ln^n(1/(1-x)) = \sum_{\ell=1} C_{n,\ell} x^\ell$. They can be calculated recursively,

$$C_{1,\ell} = \frac{1}{\ell} \qquad C_{n,\ell} = \sum_{p=1}^{\ell-1} \frac{C_{n-1,\ell-p}}{p}. \qquad (A.2)$$

Note that $C_{n,l} = 0$ for $l < n$, $C_{n,n} = 1$ and $C_{n,n+1} = \frac{n}{2}$. To close the sequence of mappings we need to express $B \mapsto D$ or rather

$$\mathbb{E}\left(\zeta^\ell\right) = \mathbb{E}\left(\left(\frac{e^x - 1}{e^x + e^y - 1}\right)^\ell\right) \tag{A.3}$$

in terms of moments, but we already know how to do this, see Eq. (53). Now we have all pieces to close the chain of mappings, for example in terms of $b_n$:

$$\hat{b}_n = \frac{\alpha K}{2} \sum_{\ell=1}^{\infty} C_{n,\ell} \left(\sum_{p,q} D_{p,q}^{(\ell)} b_p b_q\right)^{K-1}. \tag{A.4}$$

The "hat" on the lhs of (A.4) is easily removed by the usual transformations from cumulants to central moments, see (47). Eq. (A.4) can be used to calculate the central moments of $B(y)$ up to any given order in $\varepsilon = 2^{-K}$. It is usefull to introduce another small quantity $Z_\ell = \mathcal{O}(\varepsilon)$ by

$$\sum_{p,q} D_{p,q}^{(\ell)} b_p b_q =: 2^{-l}\left(1 + Z_l\right). \tag{A.5}$$

This allows us to write Eq. (A.4) as

$$\hat{b}_n = \hat{\alpha} K \sum_{\ell=n}^{r+1} (2\varepsilon)^{\ell-1} C_{n,l} \sum_{m=0}^{r+1-\ell} \binom{K-1}{m} Z_\ell^m + \mathcal{O}(\varepsilon^{r+1}) \tag{A.6}$$

Note that $Z_\ell$ must be known up to (and including) terms of order $\varepsilon^{r+1-\ell}$ in Eq. (A.6). To find $\hat{b}_1 = \mathbb{E}(y)$ up to order $\varepsilon^2$ we need

$$Z_1 = -\frac{\delta}{2} - \frac{\delta^2}{4} + \mathcal{O}(\varepsilon^3) \qquad Z_2 = -\delta + \hat{\alpha} K \varepsilon + \mathcal{O}(\varepsilon^2) \tag{A.7}$$

with $\delta = e^{-\mathbb{E}(y)} = \mathcal{O}(\varepsilon)$. Inserting this into Eq. (A.6) we get

$$\mathbb{E}(y) = \hat{\alpha} K \left(1 - \frac{K-1}{2}\delta + \varepsilon + \frac{(K-1)(K-4)}{8}\delta^2 - (K-1)\delta\varepsilon \right. \\ \left. + \hat{\alpha} K(K-1)\varepsilon^2 + \frac{4}{3}\varepsilon^2\right) + \mathcal{O}(\varepsilon^3) \tag{A.8}$$

A series expansion in $\varepsilon$ requires $\delta$ to be expressed in terms of $\varepsilon$. For that we need to know $\hat{\alpha}$ in terms of $\varepsilon$ (this is where self-consistency sneaks in), but luckily we need to know $\hat{\alpha}$ only up to first order in $\varepsilon$, so we can use (64) to get

$$\delta = \varepsilon + \left(\frac{K(K-2)}{2}\ln(2) + \frac{K}{2}\right)\varepsilon^2 + \mathcal{O}(\varepsilon^3). \tag{A.9}$$

Note that this expansion strictly holds only at $\hat{\alpha} = \hat{\alpha}_c$, the condition that underlies (64). Inserting this into the series for $\mathbb{E}(y)$ provides us with

$$\frac{\mathbb{E}(y)}{\hat{\alpha}K} = 1 + \frac{3-K}{2}\varepsilon$$
$$+ \left(\frac{17}{6} - \left[\frac{11}{8} + \frac{3}{2}\ln(2)\right]K - \left[\frac{1}{8} - \frac{7}{4}\ln(2)\right]K^2 - \frac{\ln 2}{4}K^3\right)\varepsilon^2 \quad (A.10)$$
$$+ \mathcal{O}(\varepsilon^3).$$

Similar calculations give

$$\frac{b_2}{\hat{\alpha}K} = 2\varepsilon + \left(6 - 2\left[1 + \ln(2)\right]K + 2\ln(2)K^2\right)\varepsilon^2 + \mathcal{O}(\varepsilon^3) \quad (A.11)$$

and

$$\frac{b_3}{\hat{\alpha}K} = 4\varepsilon^2 + \mathcal{O}(\varepsilon^3). \quad (A.12)$$

For the series expansion of the complexity we need to consider the $\Sigma_0$, Eq. (15) and the integrals $I_s$, Eq. (13). The Taylor coefficients

$$L_{p,q} = \frac{1}{p!\,q!} \frac{\partial^{p+q}}{\partial^p y\, \partial^q z} \ln(e^y + e^z - 1)\bigg|_{y=z=\mathbb{E}(y)}. \quad (A.13)$$

$$L_{0,0} = \ln(2e^{\mathbb{E}(y)} - 1) = \ln(2) + \mathbb{E}(y) - \frac{\delta}{2} - \frac{\delta^2}{8} + \mathcal{O}(\varepsilon^3) \quad (A.14)$$

$$L_{0,2} + L_{2,0} = \frac{1-\delta}{(2-\delta)^2} = \frac{1}{4} + \mathcal{O}(\varepsilon^2) \qquad L_{2,2} = -\frac{2+\delta^2}{4(2-\delta)^4} = -\frac{1}{32} + \mathcal{O}(\varepsilon) \quad (A.15)$$

$$L_{0,3} + L_{3,0} = -\frac{(1-\delta)\delta}{3(2-\delta)^2} = \mathcal{O}(\varepsilon) \qquad L_{0,4} + L_{4,0} = -\frac{1}{96} + \mathcal{O}(\varepsilon^2) \quad (A.16)$$

allow us to write

$$\Sigma_0 = \sum_{p,q} L_{p,q}\, b_p\, b_q = L_{0,0} + (L_{0,2} + L_{2,0})b_2 + L_{2,2}b_2^2 + (L_{0,3} + L_{3,0})b_3 + \mathcal{O}(\varepsilon^3)$$
$$= \ln(2) + \mathbb{E}(y) - \frac{\delta}{2} - \frac{\delta^2}{8} + \frac{b_2}{4} - \frac{b_2^2}{32} - \frac{b_4}{96} + \mathcal{O}(\varepsilon^3). \quad (A.17)$$

The second contribution to the complexity can be expressed in terms of the small quantities $Z_\ell$ defined in (A.5),

$$\alpha\Big[K\,I_{K-1} - (K-1)\,I_K\Big] = \hat{\alpha} \sum_{\ell=1}^{\infty} \frac{\varepsilon^{\ell-1}}{\ell} \sum_{m=0}^{K} \Big[(K-1) - 2^\ell(K-m)\Big] \binom{K}{m} Z_\ell^m$$
$$= \hat{\alpha} \sum_{\ell=1}^{3} \frac{\varepsilon^{\ell-1}}{\ell} \sum_{m=0}^{3-\ell} \Big[(K-1) - 2^\ell(K-m)\Big] \binom{K}{m} Z_\ell^m + \mathcal{O}(\varepsilon^3),$$
(A.18)

and $Z_1$ and $Z_2$ are given in (A.7). Putting together all pieces we arrive at the series expansion of the complexity $\Sigma$ up to second order in $\varepsilon$,

$$\Sigma(\alpha) = \ln(2) - \hat{\alpha} - \frac{1 + \ln(2)}{2}\varepsilon$$
$$+ \left(\frac{1}{8} - \frac{\ln(2)}{12} + \frac{3\ln(2) - 2}{8}K - \frac{\ln(2)[1 + 2\ln(2)]}{8}K^2\right)\varepsilon^2 + \mathcal{O}(\varepsilon^3).$$
(A.19)

Solving for $\hat{\alpha}$ gives the expansion for $\hat{\alpha}$ up to second order. Obviously this approach can be extended to higher orders, although the computations get more and more involved. We used a computer algebra package (Maple) to derive the next orders of the series expansion

$$\hat{\alpha}_c(K) = \ln(2) + \hat{\alpha}_1 \varepsilon + \hat{\alpha}_2 \varepsilon^2 + \hat{\alpha}_3 \varepsilon^3 + \dots.$$
(A.20)

The $\hat{\alpha}_i$ are polynomials in $K$ with coefficients that are rational polynomials in $\ln(2)$. The coefficients $\hat{\alpha}_1, \hat{\alpha}_2, \hat{\alpha}_3$ are given below. Higher order coefficients up to $\hat{\alpha}_7$ are available upon request from the authors. The quality of the expansion up to seventh order can be seen in Fig. 8 and Table 1. Note that there exist also nonanalytic terms in $\varepsilon$, because we dropped some corrections of order $\tau$ which in turn behaves as $\varepsilon^{1/(2\varepsilon)}$.

$$\hat{\alpha}_1 = -\frac{1 + \ln(2)}{2}$$
(A.21)

$$\hat{\alpha}_2 = \frac{1}{8} - \frac{\ln(2)}{12} + \frac{3\ln(2) - 2}{8}K - \frac{\ln(2) + 2\ln^2(2)}{8}K^2$$
(A.22)

$$\hat{\alpha}_3 = \frac{1}{16} - \frac{\ln(2)}{24} + \left(\frac{3\ln(2) - 2}{8}\right)K - \left(\frac{13\ln^2(2) - 3\ln(2) + 1}{8}\right)K^2 +$$
$$\left(\frac{14\ln^3(2) + 15\ln^2(2) - 4\ln(2)}{24}\right)K^3 - \left(\frac{4\ln^3(2) + \ln^2(2)}{16}\right)K^4.$$
(A.23)

## Acknowledgements

The large scale simulations to find $\alpha_c$ with the population dynamics algorithms have been done on the Beowulf cluster TINA[34]. This work has been supported in part by the European Community's Human Potential Programme under contract HPRN-CT-2002-00319, STIPCO and by the German science council (DFG) under grant ME2044/1-1. S.M. and M.M. enjoyed the hospitality of the ICTP where part of this work was done.